\documentstyle[aaspp4,11pt]{article}
\begin{document}
\title{Cosmological Luminosity Evolution of QSO/AGN Population}
\author{Yunyoung Choi$^{1,2}$, Jongmann Yang$^{1,2}$, and Insu Yi$^{1,2,3,4}$}
\affil{$^1$Center for High Energy Astrophysics and Isotope Studies,
Research Institute for Basic Sciences; yychoi@astro.ewha.ac.kr, 
yi@astro.ewha.ac.kr}
\affil{$^2$Department of Physics, Ewha University, Seoul 120-750, Korea;
jyang@astro.ewha.ac.kr}
\affil{$^3$Korea Institute for Advanced Study, Seoul, Korea}
\affil{$^4$Institute for Advanced Study, Princeton, NJ 08540}

\begin{abstract}

We apply the observed optical/X-ray spectral states of the Galactic black hole 
candidates (GBHCs) to the cosmological QSO luminosity evolution under the
assumptions that QSOs and GBHCs are powered by similar accretion processes
and that their emission mechanisms are also similar. 
The QSO luminosity function 
(LF) evolution in various energy bands is strongly affected by the spectral 
evolution which is tightly correlated with the luminosity evolution. 
We generate a random sample of QSOs born nearly synchronously by allowing
the QSOs to have redshifts in a narrow range around an initial high redshift,
black hole masses according to a power-law, and mass accretion rates near
Eddington rates. The QSOs evolve as a single long-lived population on the
cosmological time scale. The pure luminosity evolution results 
in distinct luminosity evolution features due to the strong spectral 
evolution. Most notably, different energy bands (optical/UV, soft X-ray, 
and hard X-ray) show different evolutionary trends and the hard X-ray LF 
in particular shows an apparent reversal of the luminosity
evolution (from decreasing to increasing luminosity) at low redshifts, which 
is not seen in the conventional pure luminosity evolution scenario without 
spectral evolution. The resulting mass function of black holes (BHs), 
which is qualitatively consistent with the observed QSO LF evolution, 
shows that QSO remnants are likely to be found as BHs with masses 
in the range $10^8-5\times 10^{10} M_{\odot}$. The long-lived single 
population of QSOs are expected to leave their remnants as supermassive BHs 
residing in rare, giant elliptical galaxies.

\end{abstract}


\section{INTRODUCTION}

QSOs first appear at high redshifts $z>3$ and their peak activities are
reached at $z\sim 2-3$ (e.g. Peterson 1997). The general evolutionary trend 
is that the number of bright QSOs rapidly decrease at $z<2$. 
Such an observationally established
fact has been interpreted in terms of the two drastically different theoretical
models. In the pure luminosity evolution (PLE) model (e.g. Mathez 1976,
Boyle et al. 1993, Mathez et al. 1996, Page et al. 1996),
a single generation of long-lived (e.g. $\sim 10^9$ yr) QSOs form at 
high z and evolve mainly by decreasing their luminosities. 
In the other model (density evolution model), 
a multiple of QSO populations rise and fall successively and conspire to 
result in the observed evolutionary trend. In the latter, the evolution
is largely determined by the QSO number density evolution (Haehnelt et al. 
1997 and references therein). In the PLE model, due to the longevity of
the QSOs, the massive BH remnants as massive as 
$\sim 10^{9-10}M_{\odot}$ are expected in a small number of massive galaxies
with quiescent nuclei. The density evolution model suggests that a massive 
BH is a normal feature of the central region of essentially every nearby 
galaxy (e.g. Magorrian et al. 1998, Salucci et al. 1998). 
It is to be further studied whether both
the luminosity evolution and the density evolution co-exist in the observed
QSO evolution (e.g. Miyaji et al. 1998). It is also unclear in the density
evolution model why QSOs forming at low $z$'s have lower luminosities
than their high $z$ counterparts. Therefore, it is still interesting to
see whether the PLE model can be further probed using some testable 
predictions.

If QSOs are powered by massive
accreting BHs, the QSO luminosities are essentially determined by
the black hole masses and their mass accretion rates
(e.g. Frank et al. 1992). In this sense, the
QSO luminosity evolution has to be related to the evolution of BH
masses or the evolution of the mass accretion rates or both. In most of the
previous studies in the context of the PLE model, the QSO luminosities have
been interpreted loosely as bolometric luminosities or some band
luminosities simply proportional to the bolometric luminosities (e.g. Yi
1996 and references therein). However, as
observed in some Galactic black hole candidates (GBHCs), luminosities are
strongly correlated with spectral states (e.g. Rutledge et al. 1998). 
That is, as luminosities vary,
emission spectra change so much that some energy bands' luminosities
no longer reflect the bolometric luminosity changes (cf. Narayan et al.
1998, Esin et al. 1998). In the case of
QSOs, direct spectral changes correlated with luminosities in an individual
QSO have not been established. 
However, given the striking similarities in accretion flows'
emission properties between QSOs (or active galactic nuclei (AGN) in general),
and GBHCs, a correlation between luminosity and spectrum in QSOs
is quite plausible. The Seyferts' X-ray spectral photon
index $\sim 1.5-1.7$ is similar to that of the hard state in GBHCs 
(Rutledge et al. 1998).
The soft X-ray emission in bright GBHCs (e.g. McClintock 1998)
is reminiscent of the so-called optical/UV/soft X-ray "big blue bumps" 
in bright active galactic nuclei (Frank et al. 1992). 
It is quite reassuring that the recently 
observed Galactic superluminal sources appear to be a scaled-down version
of the superluminal radio sources in extragalactic nuclei 
(Mirabel and Rodriguez 1998). Therefore, we assume that the QSO luminosity
evolution is accompanied by a strong spectral evolution.

The main effect of the spectral evolution is that the luminosity evolution
in different energy bands differ significantly.
We therefore investigate the luminosity 
evolution trend in various energy bands (Choi et al. 1999). 
Our approach can eliminate some potentially
serious uncertainties in the QSO luminosity evolution models
when only the bolometric luminosity evolution is considered (cf. Yi 1996).
In this paper, we attempt to explain the cosmological evolution of QSOs
using a model in which the LF of the QSO population is constructed from one 
generation of long-lived QSOs with an evolutionary time-scale of 
$\sim 5 \times 10^{9}$yr for a flat universe with no cosmological constant
(Yi 1996). 

\section{Luminosity Function Evolution}

We consider a single QSO population in terms of the LF which evolves in $z$.
Each QSO is turned on at an initial $z$ with an initial BH mass $M$ and an
initial mass accretion rate ${\dot M}$. An ensemble of QSOs evolve 
individually while their luminosities and spectra evolve in a correlated manner
(Choi et al. 1999), which is directly reflected in the LF evolution.

Our approach to the QSO evolution problem hinges on the hypothesis
that the luminosity-correlated QSO emission spectra are
physically similar to those of the GBHCs which show spectral transitions
closely connected to luminosity levels (e.g. Rutledge et al. 1998,
Narayan et al. 1998 and references therein). 
For instance, optical/UV/soft X-ray emission
from geometrically thin, optically thick accretion disks around supermassive
BHs contributes to the big blue bump postulated in QSOs
(Frank et al. 1992). Based on the X-ray emission spectra
observed in the GBHCs, QSOs spectral states are assumed to be directly
related to luminosities. In this picture, hard X-ray emission generally
requires a scattering corona with energetic electrons or advection-dominated
accretion flows (ADAFs) (e.g. Zdziarski 1998).
Our main assumptions are as follows.
(i) The QSOs are powered by massive accreting BHs with accretion flows.
Therefore, emission properties directly reflect the underlying accretion
flow structure which is mainly determined by the mass accretion rate.
We take the dimensionless accretion rate, 
$\dot{m} \equiv \dot{M}/ \dot{M}_{Edd}$,
which is defined as a physical accretion rate scaled by the mass-dependent
Eddington accretion rate $\dot{M}_{Edd}=1.4\times 10^{26} M_8 g/s$
where $M_8=M/10^8M_{\odot}$ and the 10\% radiative efficiency has been assumed.
(ii) The QSO emission is largely composed of three spectral states,
defined by the X-ray spectral hardness;
"Soft/High" state (HS), "Hard/Low" state (LS), in addition,
an "Off" state (OS which is included in LS), and "Very High" state (VHS)
deduced from the observed spectral states of GBHCs (Rutledge et al. 1998).
Although direct evidence for QSO spectral changes in a single QSO does not 
exist (e.g. in part due to long physical time scales appropriate for QSOs
and observational flux limits),
it is plausible that QSOs have undergone considerable luminosity and
spectral changes in the cosmological context.

We explain the cosmological evolution of QSOs using the spectral states
discussed above, which are mainly determined by the dimensionless accretion
rate, $\dot{m}$ (Yi 1996, Narayan et al. 1998).
(i) At high accretion rates, $\dot m >1$, the spectral state
of QSO corresponds to VHS. While HS, LS, and OS are well recorded in the
GBHCs, the existence and the exact nature of VHS is less clear 
(e.g. Narayan et al. 1998). We assume that the X-rays
emitted during VHS originate in the scattering hot corona which is thought to
exist above the disk at high accretion rates. The main accretion flow itself 
contributes to thermal soft X-ray/UV/optical emission. We adopt a fraction 
$\sim 10\%$ of the total luminosity for the X-ray luminosity
while the bolometric luminosity, $L=\eta \dot{M} {c^2}$ 
(with $\eta \approx 0.1$, i.e. a high efficiency) is assumed.
We identify the X-ray spectrum of this state with a power law photon index 
$\Gamma \approx 3.2$ (Rutledge et al 1998) and the optical/UV continuum with 
the thermal disk emission determined by a set of $M$ and ${\dot M}$ 
(Frank et al. 1992). 
However, since VHS is poorly understood,
the X-ray and bolometric efficiencies remain largely unconstrained.
This uncertainty is significant for the QSO evolution near the initial
birth epoch near birth. Theoretically, the accretion flow could be in the
form of the so-called slim disk with super-Eddington accretion rates
(Szuszkiewicz et al. 1996).
(ii) At lower accretion rates, $0.01\leq \dot{m}\leq 1$, the
accretion flow corresponds to the geometrically thin disk radiating optically
thick thermal emission (e.g. Frank et al. 1992) with a high efficiency 
($\eta \approx 0.1$).
The corresponding spectral state is HS. X-ray luminosity of this state is
also calculated in the same manner as in VHS (i.e. 10{\%} efficiency for
the bolometric luminosity, $L=\eta \dot{M} {c^2}$ with $\eta \approx 0.1$).
The X-ray spectrum is identified with a single power law of photon index, 
$\Gamma \approx 6.0$, deduced from the observed HS of the GBHCs
(Rutledge et al. 1998).
(iii) As the accretion fuel is gradually exhausted, the accretion rate falls
below the critical rate for the advection-dominated accretion flows (ADAFs), 
$\dot{m} < 0.01$. The corresponding spectral state is LS.
When the accretion rate falls further, the OS, which is seen in nearby,
X-ray bright galactic nuclei appears (Yi \& Boughn 1998, 1999 and references
therein). We assume that the OS is a direct
extension of LS to lower luminosities for $\dot{m}< 0.001$. 
During LS/OS, the bolometric luminosity is adequately described
by $L \approx 30 \dot{m}^x L_{Edd}$ with $ x \approx 2$ 
(e.g. Narayan \& Yi 1995, Yi 1996, Yi \& Boughn 1998, 1999). 
In this low efficiency regime, the accretion flow becomes optically thin and
radiates very weakly in optical/UV. The nonthermal emission
dominates from optical/UV to X-rays (Narayan \& Yi 1995). 
The spectral energy distribution
roughly follows a power law with photon index $\Gamma \approx 1.7$ which
is observed in GBHCs in LS. The simplified spectral shapes of the 
above-discussed states are displayed in Figure 2(a). The similar spectral
shapes for the GBHCs have been applied to several systems 
(Narayan et al. 1998, Esin et al. 1998).

We consider a population of QSOs which are born randomly within the prescribed
parameter range. 
The initial birth occurs within such a narrow range of redshifts
that the subsequent QSO evolution is largely synchronous.
(i) First, we generate a QSO population which is large enough to avoid
the small-number statistics problems. The total number of QSOs in the sample
is $10^4$. The LF evolution is obtained under the assumption that
QSOs evolve as a long-lived single population with the cosmological evolution
time scale, $t_{evol}\simeq 5 \times 10^{9}$yr (Yi 1996).
QSOs' initial birth redshifts are randomly distributed following a Gaussian 
distribution with its center at $z=4$ and width of 1.
The initial BH masses are also randomly chosen from an initial MF 
which is defined as a single power-law slope of 2.5 in the range 
$10^6 \sim 10^9 M_{\odot}$. The BHs are initially accreting at randomly
chosen accretion rates around their respective mass-dependent Eddington 
rates, $\dot{m} \sim 1$. The accretion rates are assumed to be distributed
as a Gaussian with its center at ${\dot m}=1$ and width of 0.1.
(ii) Each QSO in the sample evolves as the mass accretion in the QSO decreases
and the BH mass grows through mass accretion starting with the initial
conditions. The evolutionary time-scale, $t_{evol}$ is taken as a fraction
of the cosmic time $t_{age}$ for a flat universe ($q_0 = 0.5$) with no
cosmological constant: $t_{evol} = 0.5 t_{age}$, $t_{age} = 2t_H /3
= 2/3H_0$, with $H_0 = 50$km s$^{-1}$/Mpc. The mass accretion rate
decreases exponentially with the characteristic e-folding time scale 
$t_{evol}$ (e.g. Yi 1996). The remarkably synchronous birth and evolution
of QSOs could indirectly support the global mass accretion evolution
(e.g. Small \& Blandford 1992, Turner 1991, and references therein for related
discussions) assumed in the present study (Yi 1996 and references therein). 
As ${\dot m}\propto {\dot M}/M$ decreases
during the course of the cosmological evolution (in part due to decreasing
${\dot M}$ and more importantly due to the growth of the BH mass),
the luminosity and spectral state changes as described above. 
(iii) We then construct the differential LFs in various energy bands at 
different $z$'s. One of our major goals is to derive and compare LFs at various 
energy bands. The LFs are shown in the redshift range, $0<z<4$, and
the luminosity range, $10^{41}<L<10^{46}$erg/s. 
The LF and its evolution contains several free parameters:
a slope of initial BH mass function, BH mass range, $t_{evol}$, and
initial $\dot m$ for a given cosmological model.
The best fit values for these parameters are chosen for qualitative and
quantitative comparisons. The evolutionary timescale $t_{evol}$ controls
the overall evolution.

The evolution of the LFs in different energy bands is shown in Figure 1.

(i) In the observable luminosity range ($>10^{42}$erg/s), the QSO LFs show an
apparent break from a simple power-law (reflecting the initial BH mass 
function). That is, even for our simple single power-law initial mass
function (and hence initial LF), the LFs at late times show that the steep
power law slope at high luminosity end turns over to a much flatter slope at
low luminosity end, which is observed in the QSO samples 
(e.g. Boyle et al. 1993, Jones et al. 1997). 
This particular feature is surprisingly in good agreement
with the trend of the  observed LFs (e.g. the combined ROSAT
and EMSS samples by Jones et al. (1997) and ROSAT samples by Miyaji et al. 
1998). Our evolution model shows that LFs may appear to deviate from
the pure luminosity evolution (in which only bolometric luminosity is
considered) even when the QSO evolution is essentially driven by
the luminosity evolution. This effect is obviously caused by the spectral
changes accompanying the luminosity changes.
We point out that this particular result implies there is no strong
evidence for cosmological evolution of the space density of
low-luminosity AGN in contradiction to Miyaji et al. (1998).

(ii) The luminosity functions in X-ray bands are differently affected by 
the spectral changes caused by the transition of accretion flows.
Most notably, the most significant spectral change occurs when the QSO spectrum
changes from HS (thin disk, $\Gamma \approx 6.0$) to LS (ADAF, $\Gamma \approx
1.7$), which to the lowest order induces a sudden luminosity decrease. 
However, since ADAFs are relatively hard X-ray bright (Yi \& Boughn
1998, 1999), QSOs' hard X-ray LFs
show a trend very different from other energy bands as shown in Figure 1
(also see below). The QSOs undergo an accretion flow transition at a
critical redshift $z_c\sim 1$, which corresponds to an epoch where
$\dot{m}=0.01$ (Narayan \& Yi 1995) as discussed above. Although the exact 
value of $z_c$
depends on various model parameters (e.g. Yi 1996), it is plausible to
identify $z_c$ with the observed sudden decline of bright QSOs.
For instance, $t_{evol} \ll \sim 5 \times 10^{9}$yr could drive
much faster evolution than the observed one. Our model $z_c$ is comfortably
close to the observed break near $z\sim 1-2$ (Hewett et al. 1993, 
Page et al. 1996).

(iii) Optical/UV ($1216 \AA, 2500 \AA, 4400 \AA$) LFs evolve much
faster than X-ray LFs at low $z$'s. At high $z>z_c$, however, optical/UV
luminosity evolution is very slow, which is in excellent agreement
with the observed trend. There also exists a brief period of evolution
during which the LF slightly moves to higher luminosity range although
such a phenomenon is hard to detect observationally.
In the case of soft X-ray LF, the slow evolution
is also seen at $z>z_c$. This is also in agreement with the results of 
Page et al. (1996).

(iv) At hard X-ray energies ($2.0-10$keV), a dramatic turnaround in the
LF evolution is identified at $z<z_c\sim 1$. That is, the
hard X-ray LF as a whole gradually shifts to lower luminosities
until $z\sim z_c$ is reached. Below $z_c$, the relatively X-ray bright
(especially in hard X-rays) ADAFs drive an increasing number of QSOs into
higher hard X-ray luminosities. This results in reversal in the direction of
the hard X-ray LF evolution as shown in Figure 1(d).
This rather surprising result is explained by the fact that QSOs have 
undergone an accretion flow transition
from a thin disk (HS) to an ADAF (LS) as ${\dot m}$ decreases.
Observationally, neither transition in hard X-ray luminosity nor
luminosity evolution itself has been clearly seen. However, once 
$2-10$keV X-ray LFs become available (e.g. Chandra),
this predicted trend should be testable.

(v) At soft X-ray energies ($0.5\sim 2.0$keV), there is no apparent 
dramatic transition.
The soft X-ray luminosity evolves roughly as $L \propto (1+z)^{k}$ 
with $k>3$, up to $z_c$. This value is similar to but larger
than those obtained by Boyle et al. (1994) and Page et al. (1996) based 
on the pure (bolometric) luminosity evolution model. The discrepancy
could be resolved if spectral evolution is taken into account.
For a quantitative comparison between model LFs and observed ones, we convert
the space density per comoving volume in the ROSAT AGN soft X-ray LF
estimated by Miyaji et al. (1998)
to QSO LF assuming $H_0=50h_{50}km/s/Mpc$, $h_{50}=1$, and $q_0 = 0.5$.
On the whole, our model and the Miyaji et al. (1998) result agree quite
well with each other. We consider this qualitative agreement as a support for
the QSO luminosity evolution models.

The MF of BHs and its evolution can be naturally derived from our evolutionary
calculation. As long as the model QSO LF agrees with the observed LF, the
resulting MF has immediate consequences on the fate of the QSO
remnants at the present epoch. Figure 2(b) shows the differential MF
of BHs which have grown during the QSO evolution.
The initial MF with a slope of power law, $s=2.5$ has been assumed.
The BH MF suggests that the QSO remnants are likely to have BH masses 
in the range $10{^8}M_{\odot}$ to $\sim 5 \times 10^{10}M_{\odot}$. On average,
remnant BH masses are bigger than the initial masses by a factor $\sim 100$.
BH remnants in the mass range, $10^8 M_{\odot} \sim 10^9 M_{\odot}$
are most likely to be found in some present-day galaxies' centers.
This finding obviously suggests that it is
difficult for our model to explain small BHs with masses in the range
$10^{5} \sim 10^{7} M_{\odot}$ which are found in nearby, typical spiral
$L^*$ (e.g. Peebles 1993) galaxies with the comoving space
density $\sim 10^{-2} h_{50}^3Mpc^{-3}$ (Magorrian et al. 1998). 
In the present single population scenario,
the integrated comoving density of very massive QSO remnants is $\sim 4\times
10^{-5}h_{50}^3Mpc^{-3}$, which suggests 
that they are likely to be found in rare, giant elliptical galaxies.
Our model inevitably points to massive BHs with masses 
$\geq 10^{9} M_{\odot}$ (e.g. Fabian \& Canizares 1988, Mahadevan 1997) 
in elliptical galaxies as QSO remnants. 
It is possible that smaller mass black holes in galactic
nuclei might have grown without experiencing the QSO phase.

\section{Discussions}

Our model is largely consistent
with the observed QSO evolution trend, which supports the possibility that
the QSO evolution is accounted for by a single, long-lived QSO population. 
The evolution of LF shows an apparent transition at a critical redshift
caused by the accretion flow transition (Yi 1996). As far as we know, there 
is no other convincing explanation for this apparent break in the QSO 
evolution. The overall agreement between the observed LF evolution and our 
model implies that spectral states of QSOs may indeed be similar to those of 
GBHCs.

The accretion flow has undergone a transition as $\dot{m}$ has declined from
$\sim 1$ to $<0.01$ while the initial BH masses have
evolved by a factor $\sim 100$. The MF of BHs indicates that QSO remnants are 
likely to be found as massive BHs with masses in the range 
$\sim 10{^8} - 5 \times 10^{10}M_{\odot}$, which are likely to be found
as massive BHs of $\geq 10^{9} M_{\odot}$ residing in very rare
elliptical galaxies.

The distinguishable evolutionary signature at hard X-ray energies
is significant and also in principle testable.
Future confirmation or rejection of our prediction by hard X-ray luminosity 
functions (e.g. by Chandra) could provide a crucial piece of information.
It remains to be seen whether an alternative scenario,
in which QSO evolution is composed of many short-lived populations,
is also affected by the luminosity-spectrum correlation we have looked into.
The QSO spectral evolution could result in some interesting consequences
for the X-ray background although the detailed analyses require rather
sophisticated spectral calculations (e.g. Yi \& Boughn 1998 and references
therein).

\acknowledgments
JY acknowledges the partial support from KRF 1998-015-D00129 and
1998-001-D00365. IY acknowledges some useful discussions with R. Narayan on 
spectral states during the early stage of this work and
support from KRF 1998-001-D00365, MOST project 98-N6-01-01-A-06, and 
Ewha Faculty Reserach Fund 1998. IY thanks SUAM foundation for its support 
during the early stage of this work performed at IAS.

\clearpage

\noindent

\vfill\eject
\clearpage

\begin{figure}
\caption{The redshift evolution of QSO luminosity function (LF) at different
energy bands. The LF evolution has been obtained under the assumption
that QSOs evolve as a single, long-lived population 
($t_{evol}\simeq 5 \times 10^{9}$yr).
The total number of randomly generated QSOs in the sample is $10^4$. 
The comoving number density is normalized to match that of 
Miyaji et al. (1998). 
This figure shows that different energy bands show distinct LF evolution 
signatures.
Especially, at hard X-ray energies, QSOs appear to experience a rather drastic 
turnaround at low z.}
\label{autonum}
\end{figure}

\begin{figure}
\caption{(a) The simplified spectral energy distributions (SEDs) of a QSO
with a black hole mass of $10^8M_{\odot}$. From top to bottom, the decreasing
${\dot m}$'s correspond to the Very High State (VHS) with ${\dot m}>1$,
the High State (HS) with ${\dot m}=0.01-1$, and the Low State and the Off State
(LS and OS) with ${\dot m}<0.01$. The spectral and luminosity evolution
is caused by the decrease of ${\dot m}$.
(b) The differential mass function (comoving number density) of BHs at various 
redshifts. BHs form at high redshifts and grow through mass accretion during
the QSO evolution. For simplicity, the initial mass function is assumed to have
a power-law slope with index 2.5. The QSO remnants are likely to have BH masses
in the range from $10^8$ to $5\times 10^{10} M_{\odot}$. The integrated comoving
space density of QSOs is $\sim 4\times 10^{-5}h_{50}^3Mpc^{-3}$.}
\end{figure}
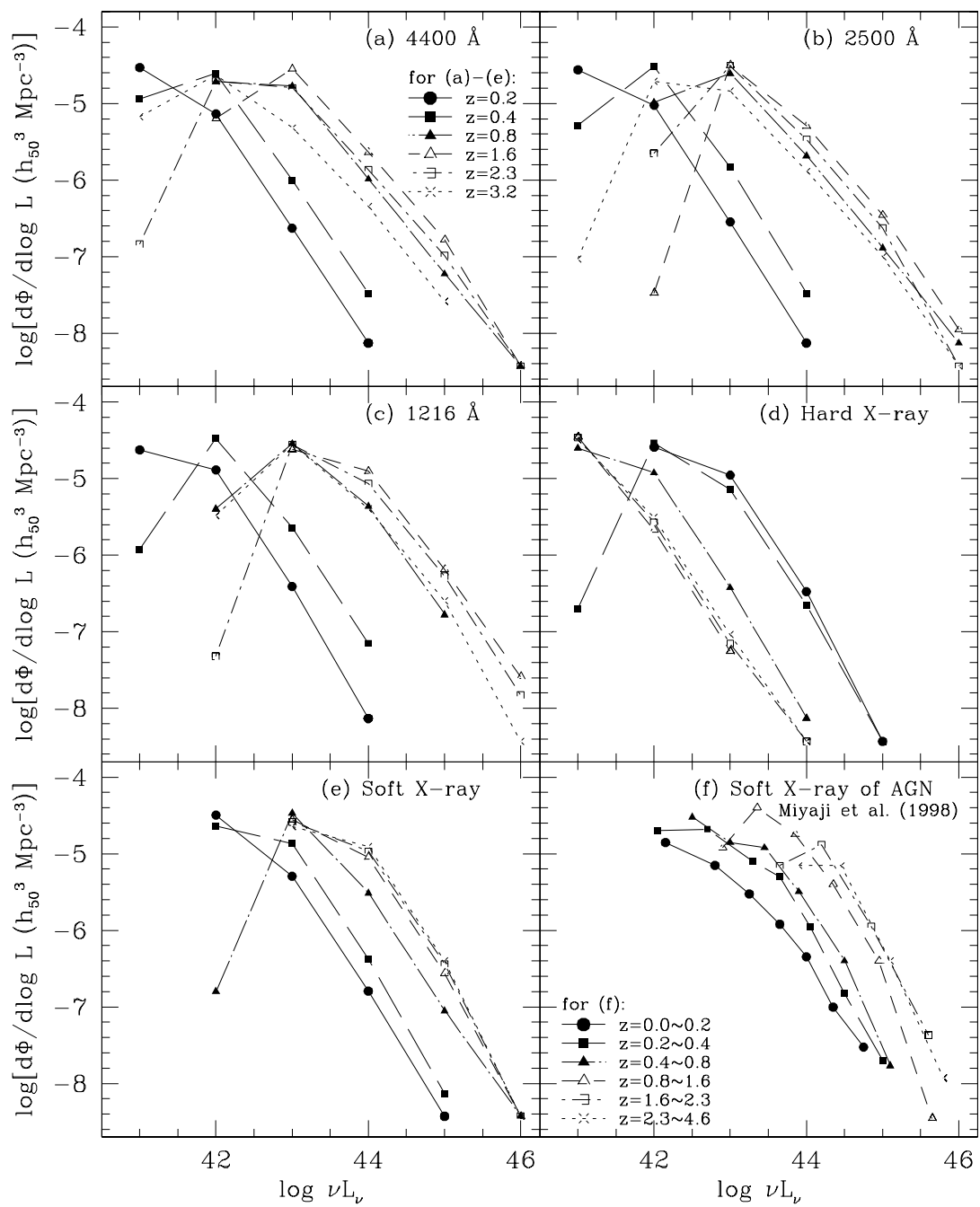
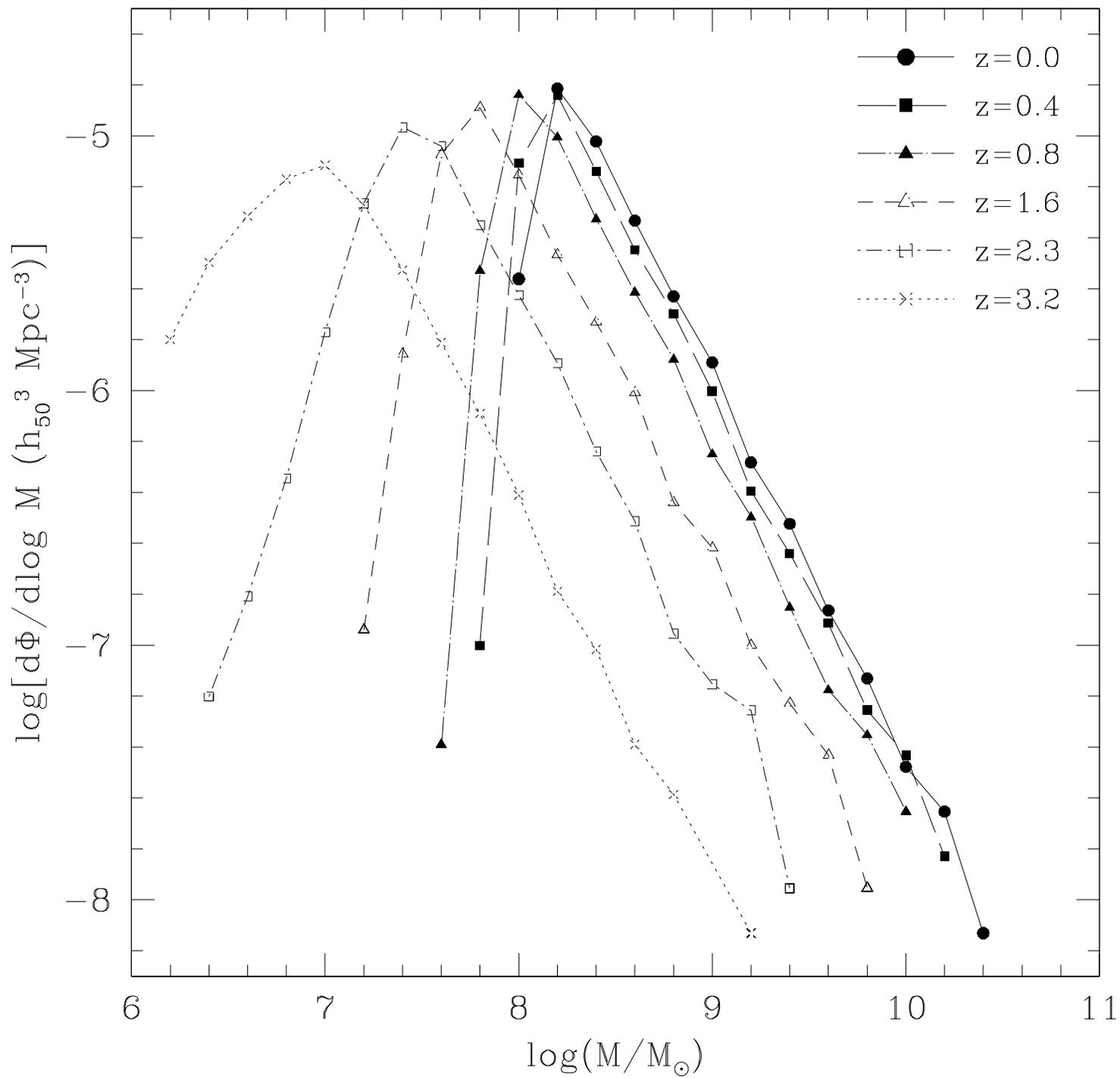
\end{document}